\documentclass[aps,prl,numerical,reprint,showpacs]{revtex4-1}  %aip,jmp     aps,prb
\usepackage[dvips]{graphicx}%
\usepackage{bm}% 
\usepackage{amsmath}
\usepackage{dcolumn}
\usepackage[normalem]{ulem} %solo sirve para tachar texto
\usepackage[colorlinks=true,linkcolor=blue,citecolor=red ,urlcolor=blue]{hyperref}

\draft
\begin{document}

%****** new commands
\newcommand{\op}[1]{{\bm{#1}}}
\newcommand{\bra}{\langle}
\newcommand{\ket}{\rangle}
\newcommand{\new}[1]{\textcolor[rgb]{0,0.3,0}{\uline{#1}}}
\newcommand{\old}[1]{\textcolor[rgb]{1,0,0}{\sout{#1}}}
\newcommand{\nota}[1]{\textcolor[rgb]{0.85,0.0,0.0}{{#1}}}
%************************

%\preprint{APS/123-QED}

\title{A triple point in 3-level systems}

\author{E. Nahmad-Achar}
\email{nahmad@nucleares.unam.mx}
\author{S. Cordero} 
%\email{sergio.cordero@nucleares.unam.mx}
%
\author{R. L\'opez-Pe\~na}
%\email{lopez@nucleares.unam.mx}
%
\author{O. Casta\~nos}
%\email{ocasta@nucleares.unam.mx}
%

\affiliation{%
Instituto de Ciencias Nucleares, Universidad Nacional Aut\'onoma de M\'exico, Apartado Postal 70-543, 04510 M\'exico DF,   Mexico  \\ }

\date{\today}

\begin{abstract}

The energy spectrum of a $3$-level atomic system in the $\Xi$-configuration is studied. This configuration presents a {\it triple point} independently of the number of atoms, which remains in the thermodynamic limit. This means that in a vicinity of this point any quantum fluctuation will drastically change the composition of the ground state of the system. We study the expectation values of the atomic population of each level, the number of photons, and the probability distribution of photons at the triple point.

\end{abstract}

%

% Photon-atom interactions, 32.80.-t

\maketitle

\section{Introduction}

The study of systems of $3$-level atoms is interesting as proposals have been made to use them as quantum memories or to manipulate quantum information, among other applications. Cavity QED systems in particular have been favored because of their advantage when subjected to coherent manipulations, and schemes have been presented for various quantum gates using 3-level atoms and trapped ions~\cite{yi, jane}.
In the presence of an electromagnetic field, few-level atomic systems present the phenomenon of {\it superradiance}~\cite{hepp} where the decay rate is proportional to the square of the number of atoms, $N_a^2$, instead of $N_a$ (the expected result for independent atom emission). There thus exists a {\it separatrix} in parameter-space which divides a so-called {\it normal} region from a region of collective behavior~\cite{cordero1, cordero2}.

\begin{figure}[!h]
\begin{center}
\includegraphics[width=0.6\linewidth]{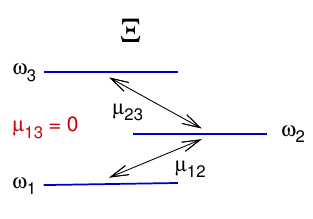}
\end{center}
\caption{Atomic $\Xi$, or {\it ladder}, configuration. The $i$-th level frequency is denoted by $\omega_i$ with the convention $\omega_1\leq\omega_2\leq\omega_3$, and the coupling parameter between levels $i$ and $j$ is $\mu_{ij}$.}
\label{f1}
\end{figure}

In this contribution we study a $3$-level atomic system in the $\Xi$-configuration (cf. Fig.~\ref{f1}) interacting through a one-mode electromagnetic field, and we show that its phase diagram for the ground state presents a triple point independently of the number of atoms, which prevails in the thermodynamic limit. We use the fidelity and the fidelity susceptibility of neighboring states to determine the sudden changes, in parameter space, in the ground state composition. We also find that, when in double resonance, there is a mirror symmetry in the energy spectrum around the energy value $E=M$, where $M$ is the total excitation number (which turns out to be a constant of motion in the rotating-wave approximation (RWA)).

%Figura 1

The intrinsic Hamiltonian for a $3$-level system in the RWA is~\cite{cordero2}
%\begin{widetext}
	\begin{eqnarray}
		\label{hamiltonian}
H = &&\ \Omega\, a^\dag a + \omega_1\,A_{11} + \omega_2\,A_{22} + \omega_3\,A_{33}  \\
- &&\frac{1}{\sqrt{N_a}} \left[\mu_{12}\left(a\,A_{21} + a^\dag\,A_{12} \right)
+ \mu_{23}\left(a\,A_{32} + a^\dag\,A_{23}\right) \right] \nonumber
	\end{eqnarray}
%\end{widetext}
where ${a}^\dag,\, {a}$ are the creation and annihilation operators of the field, ${A}_{ij}$ are the atomic operators, the $i$-th level frequency is denoted by $\omega_i$ with the convention $\omega_1\leq\omega_2\leq\omega_3$, and the coupling parameter between levels $i$ and $j$ is $\mu_{ij}$. The detuning between levels $i$ and $j$ and the field is denoted by $\Delta_{ij} = \omega_i - \omega_j - \Omega$, where $\Omega$ is the field frequency. It has $2$ constants of motion, viz., the total number of atoms $N_a = \sum_{i=1}^3 A_{ii}$, and the total number of excitations $M = a^\dag a + A_{22} + 2\, A_{33}$. As the system is non-integrable, one may solve via numerical diagonalization. A natural basis in which we diagonalize our Hamiltonian is $\vert\nu;\,q,\,r\rangle$~\cite{cordero1}. Here, $\nu$ represents the number of photons of the Fock state, $r,\, q-r$ and $N_a-q$ are the atomic population of levels $1,\, 2,\, 3$, respectively. Notice that the Hamiltonian~(\ref{hamiltonian}) is invariant under the transformation $a\to -a$ and $a^\dagger\to -a^\dagger$, which preserves the commutation relations of the bosonic operators. For this reason we consider only positive values for $\mu_{ij}$.

\section{Normal and collective regimes of the Hamiltonian system}

\begin{figure}
\begin{center}
a)\ \includegraphics[width=0.85\linewidth]{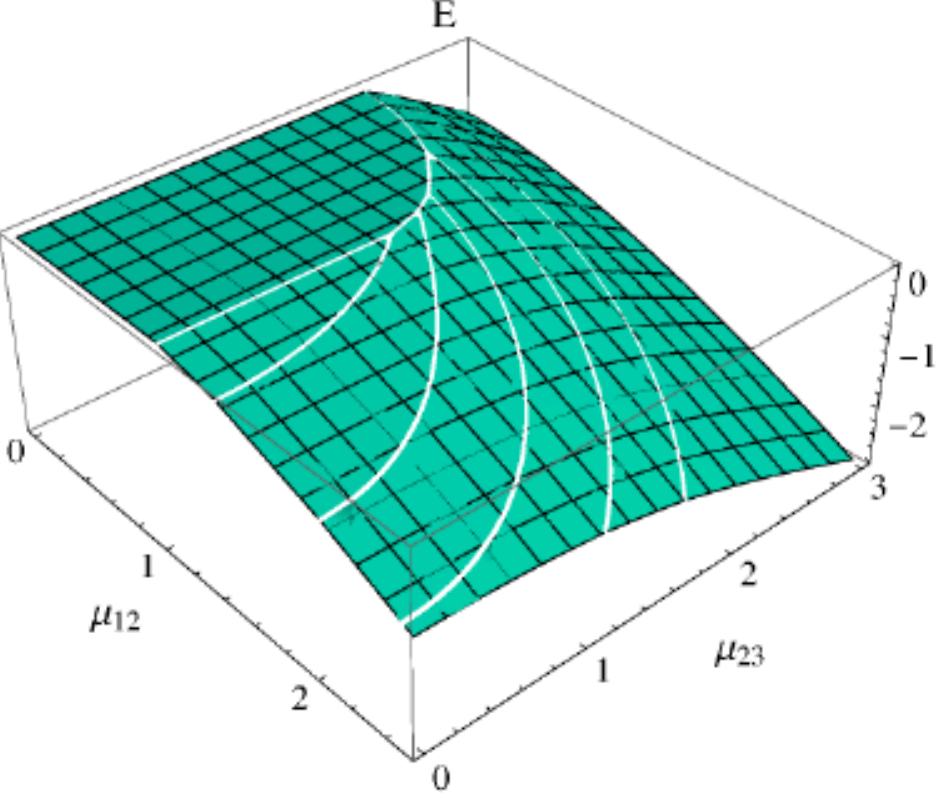} \\
b)\ \includegraphics[width=0.85\linewidth]{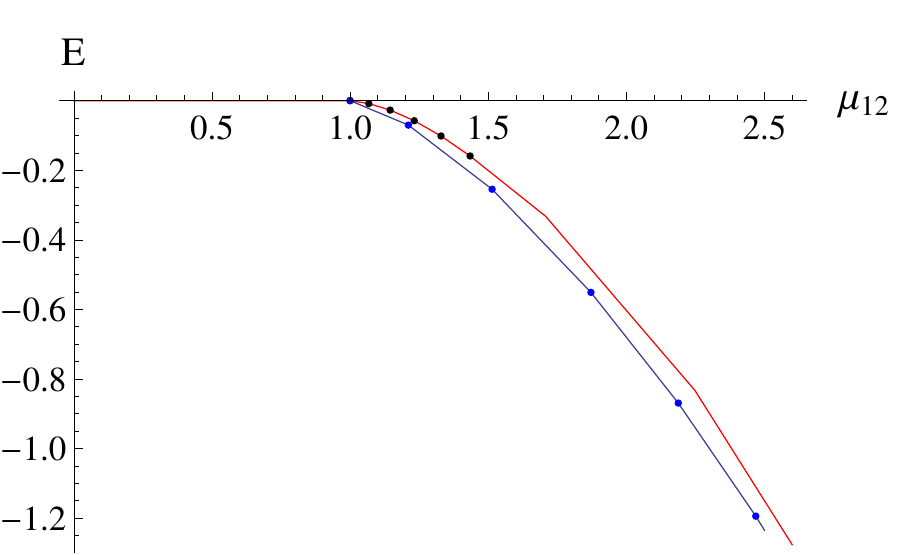} 
\end{center}
\caption{(Color online.)  a) Energy of the ground state as a function of the coupling parameters, in double resonance for $N_a=2$. The intersection lines between $M=k$ with $M=k+1$ from $k=0, \dots 5$ are also shown. At $(\mu_{12},\,\mu_{23}) = (1,\,\sqrt{2})$ the regions $M=0,\, M=1$ and $M=2$ meet: a true triple point. b) A cut at $\mu_{23}=0$ is shown for $N_a=3$ (lower, blue line) and $N_a=8$ (upper, red line).}
\label{f2}
\end{figure}

The energy of the ground state as a function of the coupling parameters $\mu_{12}$ and $\mu_{23}$ is displayed in Fig.~\ref{f2}a, in double resonance. We use $N_a=2$ for clarity of the figure. The intersection lines between $M=k$ with $M=k+1$ from $k=0, \dots, 5$ are also shown. All these {\it separatrices} may be determined by using the fidelity $F$ or the fidelity susceptibility $\chi$ of neighboring states~\cite{SimInNat} (for a complete review of the use of these concepts see~\cite{gu}), defined by
\begin{eqnarray}
	&&F(\lambda,\,\lambda+\delta\lambda) = \vert\langle\psi(\lambda)\vert\psi(\lambda+\delta\lambda)\rangle\vert^2\,, \nonumber \\[0.2in]
	&&\chi = 2\,\frac{1-F(\lambda,\,\lambda+\delta\lambda)}{(\delta\lambda)^2}.
\end{eqnarray}

%Figura 2

\begin{figure}[!h]
\begin{center}
a)\ \includegraphics[width=0.9\linewidth]{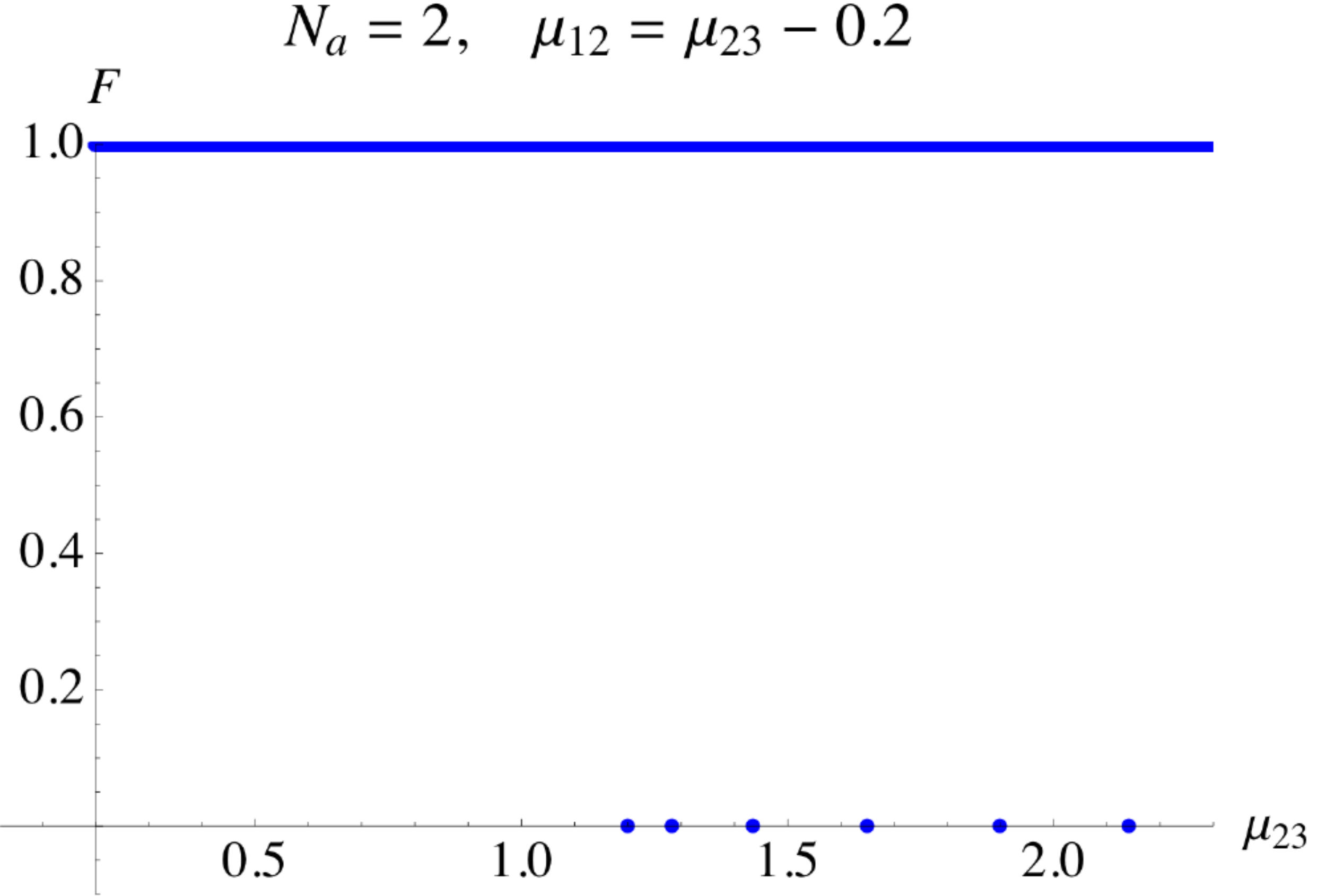}
\\[1cm]
b)\ \includegraphics[width=0.9\linewidth]{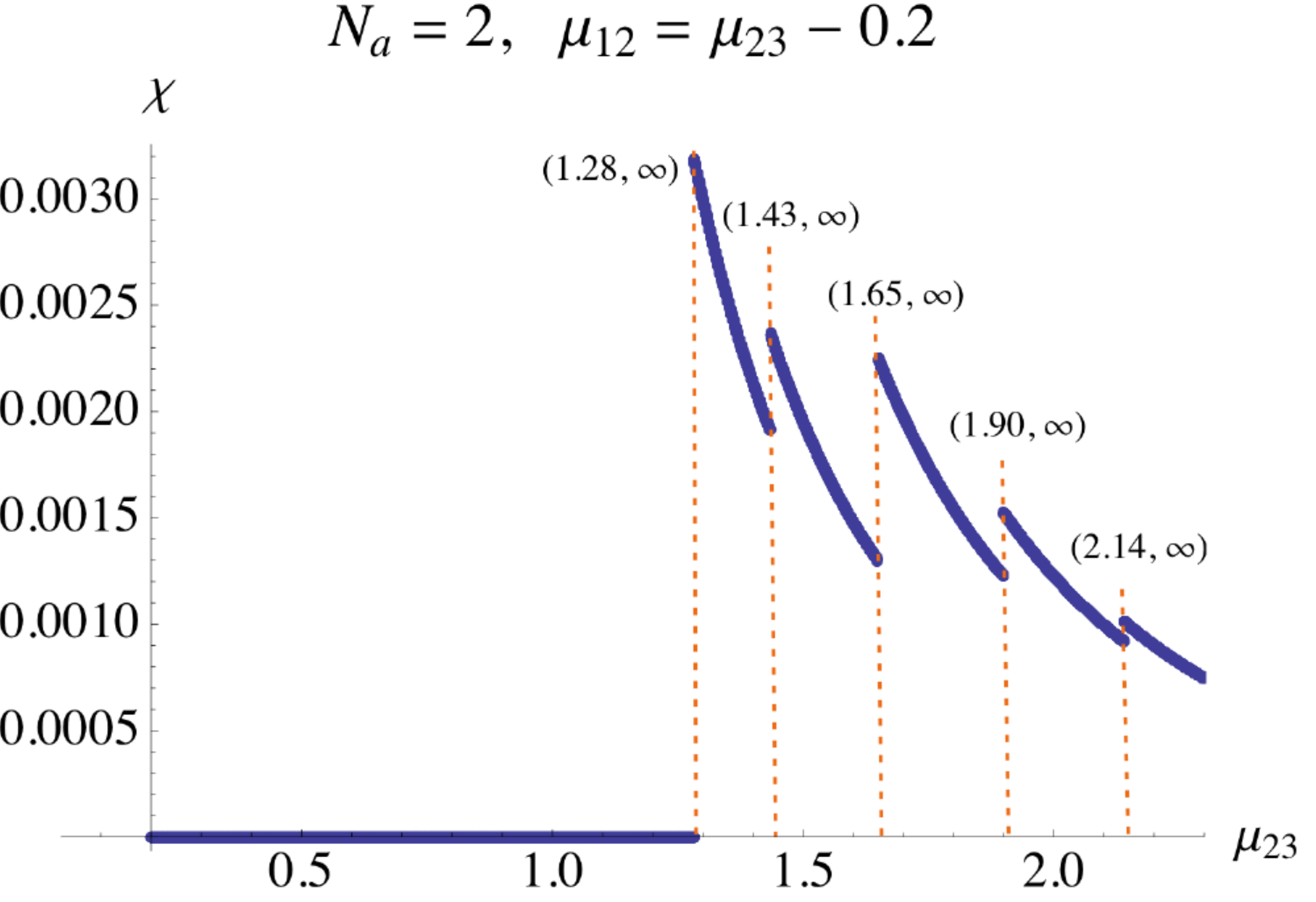}
\end{center}
\caption{a) Fidelity measure $F$ between neighboring states, along the line $\mu_{12}=\mu_{23}-0.2$ in parameter space, plotted as a function of $\mu_{23}$. It is zero at the transition points. b) Fidelity susceptibility $\chi$ as a function of the coupling parameter $\mu_{23}$; it diverges at each point where the excitation number $M$ changes.}
\label{f3}
\end{figure}

The fidelity is a measure of the distance between states, for a pure state $\vert\psi(\lambda)\rangle$ which varies as a function of a control parameter $\lambda$. The fidelity susceptibility, essentially its second derivative with respect to the control parameter, is a more sensitive quantity. Figures~\ref{f3}a,b show the zeroes attained by the fidelity measure and the divergences in the fidelity susceptibility at each separatrix, where the value of the excitation number $M$ changes. A cut along the straight line $\mu_{12}=\mu_{23}-0.2$ in parameter space was taken for good comparison with Fig.~\ref{f2}a. Here, $N_a=2$ as well. (Note that a line of the form $\mu_{12}=c\,\mu_{23}$ is not a convenient choice, as the parameter would factorize in the interaction Hamiltonian and the eigenstates would remain the same as $\mu_{12}$ changes within a region.) Whereas the fidelity measure clearly marks the transition points, it fails at distinguishing states within a region of constant $M$; the fidelity susceptibility, being a much more sensitive function, does distinguish them.

%Figura 3

The innermost {\it separatrix} divides the space into a {\it normal} region, where the number of excitations $M$ vanishes, and a {\it collective} region. This separatrix prevails in the thermodynamic limit $N_a\rightarrow\infty$. At $(\mu_{12},\,\mu_{23}) = (1,\,\sqrt{2})$ the regions $M=0,\, M=1$ and $M=2$ meet: a {\it triple point} which is fixed for all values of the number of atoms $N_a$. All other triple points as well as all other separatrices slide towards the boundary between $M=0$ and $M=1$ as $N_a$ increases, and new transition regions intersecting the region $M=0$ appear. This holds true all the way to the thermodynamic limit, where all separatrices coalesce into one. A manifestation of this phenomenon is shown in Fig.~\ref{f2}b, where a cut at $\mu_{23}=0$ of the ground state energy is shown for $N_a=3$ and $N_a=8$; the dots along the lines show where the transitions from one $M$ to the next take place along the $\mu_{12}$-axis. As $N_a$ increases, the intersection points tend to $\mu_{12}=1$. The same holds true for any other value of $\mu_{23}$.

\begin{figure}
\begin{center}
\includegraphics[width=1\linewidth]{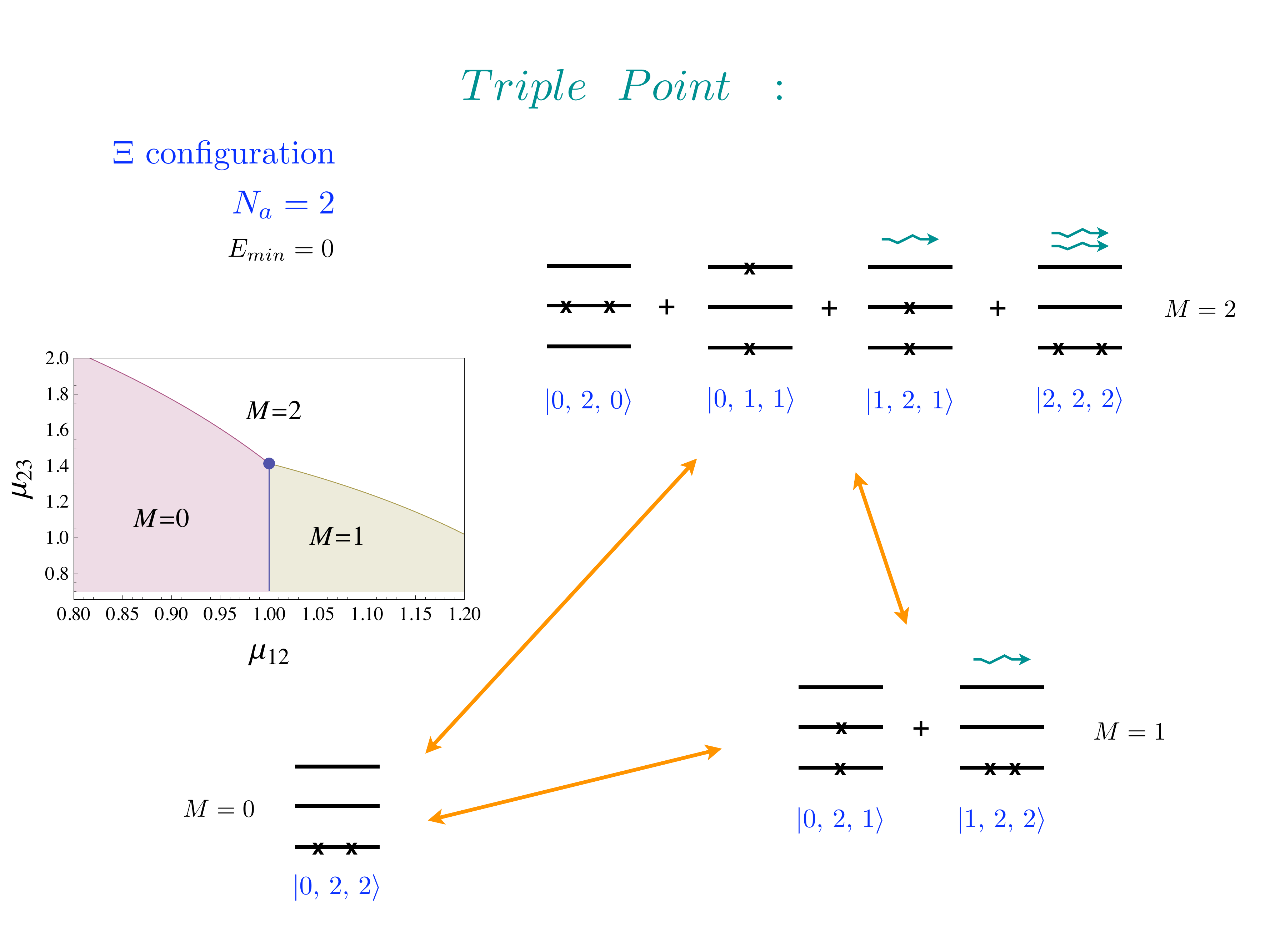} 
\end{center}
\caption{(Color online.) Triple point present at $(\mu_{12},\,\mu_{23})=(1,\,\sqrt{2})$. The figure shows the transformation of the ground state near the vicinity of this point due to quantum fluctuations, for $N_a=2$. The triple point persists for {\it any} $N_a$, including the thermodynamic limit. (Wavy lines represent photons and an ``$\times$'' represents a level occupation.)}
\label{f4}
\end{figure}

The triple point itself is shown in Fig.~\ref{f4}, together with a diagramatic description of the ground states corresponding to each value of $M$, for $N_a=2$. This means that in a vicinity of this point any quantum fluctuation will drastically change the composition of the ground state. Since the dimension of the Hilbert space does not change for larger values of $N_a$, the same composition as that shown in the figure is obtained for a larger number of atoms (in fact, since the number of excitations is low, increasing $N_a$ amounts to increase the occupation of the first level).

It is worth stressing that the existence of this triple point {\it independent} of $N_a$ is a characteristic of the $\Xi$-configuration; it does not appear in the $\Lambda$ or the $V$ configurations.

%Figura 4

\section{Properties at the triple point}

\begin{figure}
\begin{center}
a)\ \includegraphics[width=0.9\linewidth]{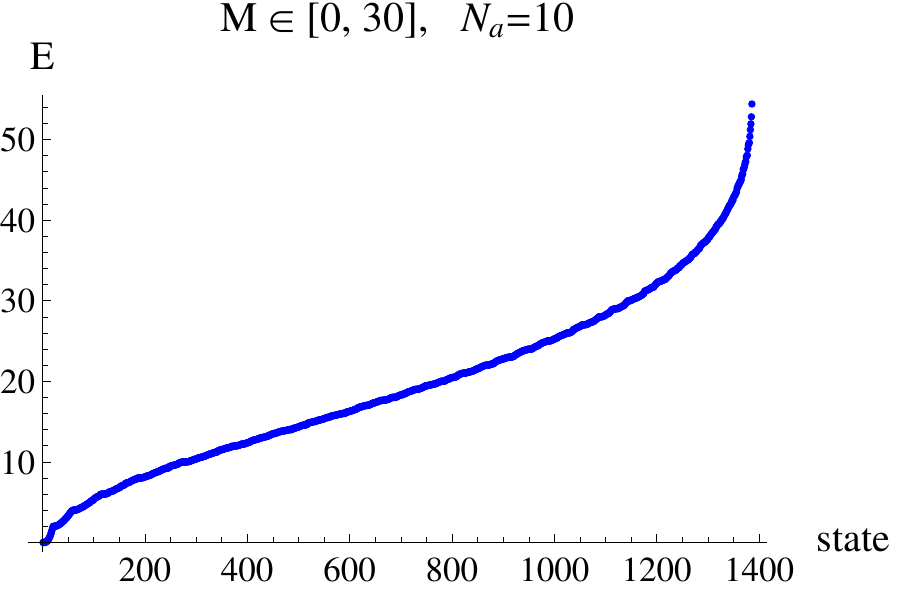} \\[1cm]
b)\ \includegraphics[width=0.9\linewidth]{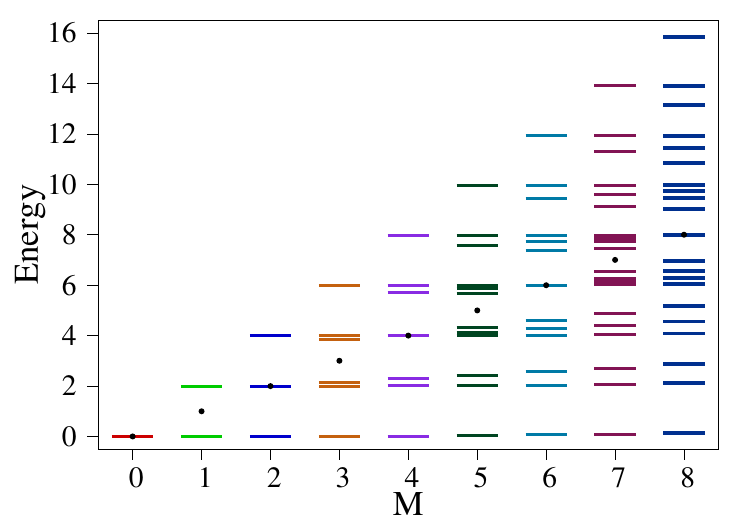} 
\end{center}
\caption{(Color online.)  Double resonant case. a) Energy spectrum for a system of $N_a=10$ particles at the triple point. The $M$ values run from $0$ to $30$.  b) Energy spectrum as function of the total excitation number for $N_a=10$ at the triple point. There is a mirror symmetry with respect to $E=M$ (indicated by dots) for each value of $M$.}
\label{f5}
\end{figure}

The energy spectrum at the triple point for a system of $N_a=10$ particles is displayed for all values of $M$ up to $30$ in Figure~\ref{f5}a, for the double resonant case. When the energy is plotted against the excitation number $M$ we observe a mirror symmetry with respect to $E=M$, and the symmetric states correspond to eigenstates whose components differ only by phases. The spectrum in Fig.~\ref{f5}b shows this for $N_a=10$. This means that all expectation values of operators will be equal for these symmetric states, except for the energy itself.

%Figura 5

The system presents degenerate states only at $E=M$, which is not always attained. For instance, if $M$ is odd and $M\leq N_a$ there are no states with $E=M$. For $M \geq 2\,N_a$ there is a degeneracy of $\lfloor N_a/2 \rfloor +1$, where $\lfloor x \rfloor$ denotes $floor(x)$. For other values it depends strongly on $M$.
	
\begin{figure}
\begin{center}
a)\ \includegraphics[width=0.9\linewidth]{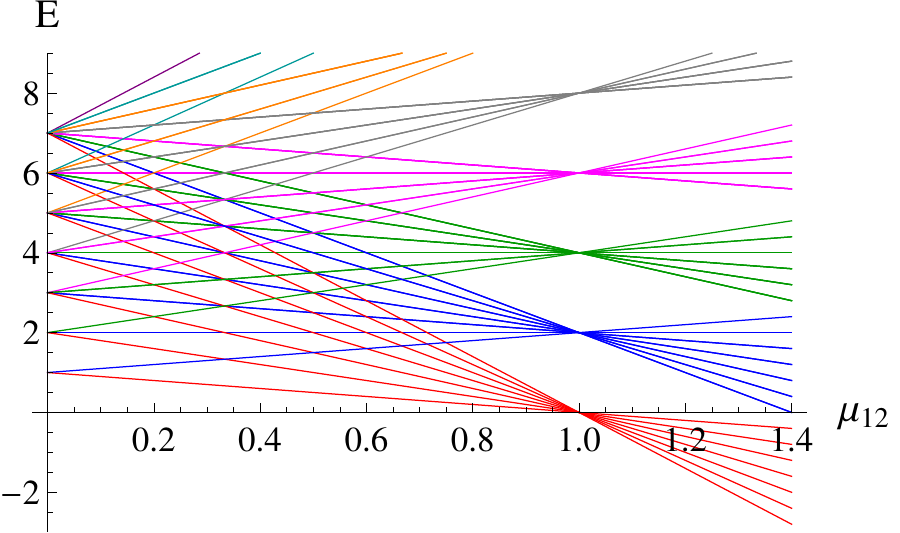}
\\[1cm]
b)\ \includegraphics[width=0.9\linewidth]{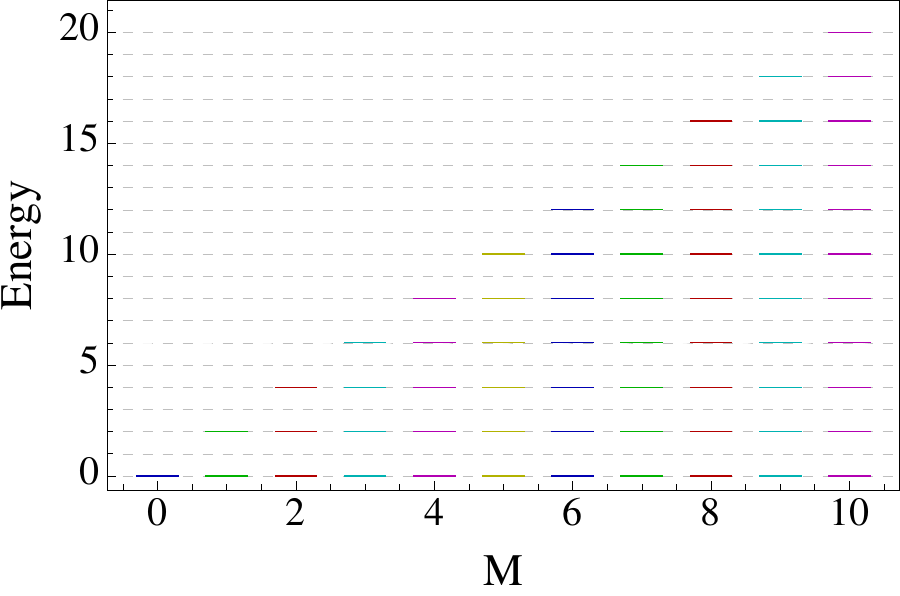} 
\end{center}
\caption{(Color online.)  a) Energy spectrum in the thermodynamic limit ($N_a \to \infty$) as a function of $\mu_{12}$ from $M=0$ to $M=7$. One can see the collapse of the successive energy levels for each value of $M$ at the energy values $E=0, 2,4,6,8$ at the triple point. b) Energy spectrum as a function of $M$ at the triple point.  Notice that for all values of $M$ we have an equidistant spectrum with only even harmonics.}
\label{f6}
\end{figure}

For small values of the total excitation number $M$ the dimension of the Hilbert space depends solely on $M$, and not on $N_a$. This allows one to study the system for a large number of atoms, including the thermodynamic limit. In this limit $N_a \to \infty$, the energy spectrum becomes independent of $\mu_{23}$ and shows a collapse of energy levels for all values of $M$ at the point $\mu_{12}= 1$. Here, the lowest energy levels for all $M$ have a value $E=0$, the next lowest energy levels for all $M$ have a value $E=2$, the next levels have $E=4$, and so on, as shown in Figure~\ref{f6}a where the energy spectrum is plotted against the coupling parameter $\mu_{12}$ for values of $M$ from $0$ to $7$. The spectrum as a function of $M$ becomes equidistant with only even harmonics in this limit (cf. Fig.~\ref{f6}b). A chain of excited-state quantum phase transitions demarcating the superradiant phase was previously reported for the Jaynes-Cummings and Dicke models of quantum optics~\cite{perez}. It was shown that the emergence of quantum chaos is caused by the precursors of the excited-state quantum phase transitions. Finite-size scaling behavior of excited-state quantum phase transitions at the mean-field level was studied for 2-level bosonic and fermionic models in~\cite{caprio}.

%Figura 6

\begin{figure}
\begin{center}
\includegraphics[width=1\linewidth]{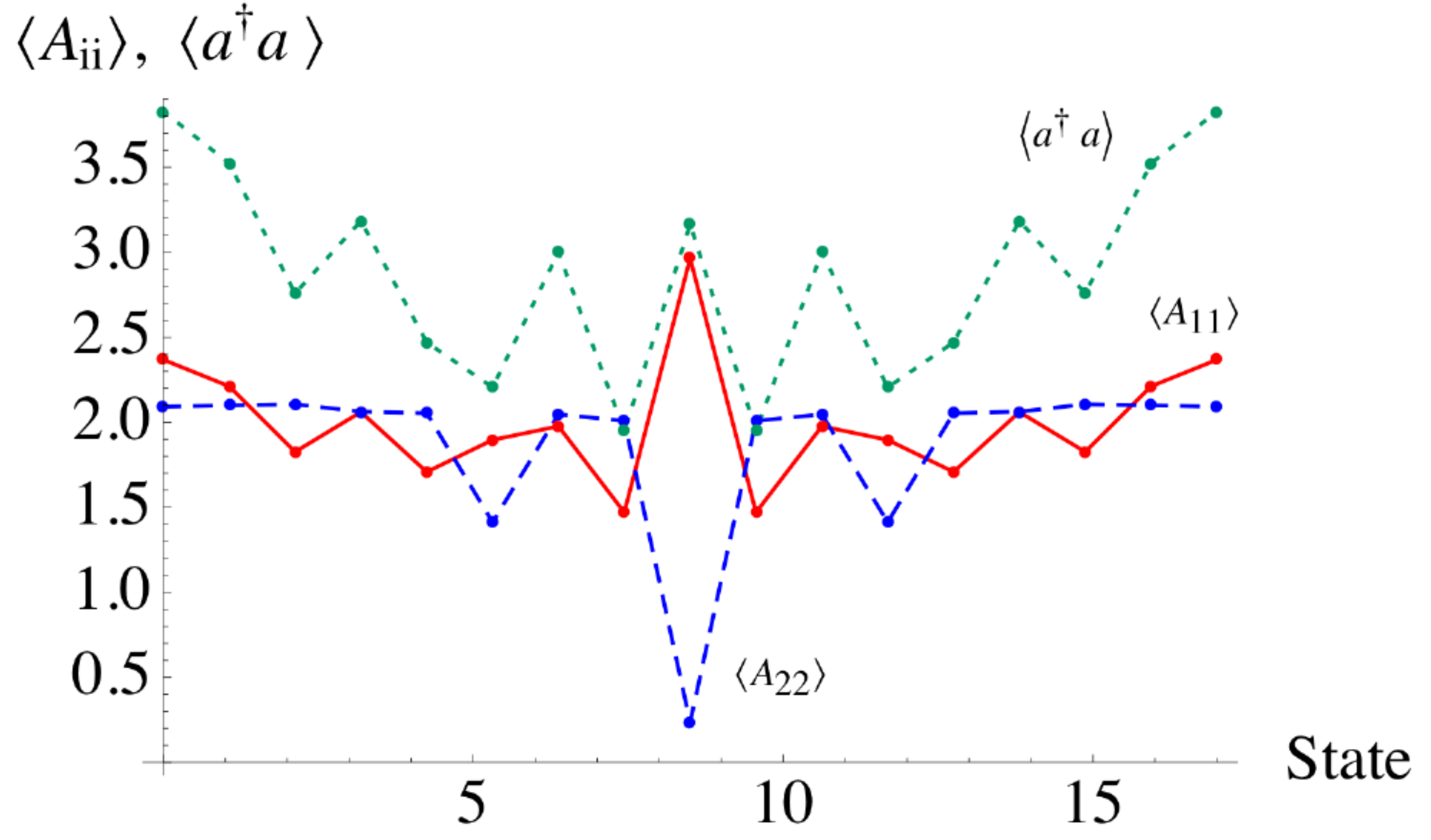}
\end{center}
\caption{(Color online.)  Expectation values at the triple point of the population operators $A_{11}$ (continuous, red), $A_{22}$ (dashed, blue), and the number of photons $\nu$ (dotted, green), for each eigenstate of the Hamiltonian. Here, $N_a=5,\,M=7$.}
\label{f7}
\end{figure}

One can also calculate the expectation values of the population operators $\langle A_{ii}\rangle_{i=1,2}$ and $\langle a^\dagger\,a\rangle$ at the triple point, for each of the eigenstates of the Hamiltonian~(\ref{hamiltonian}) Figure~\ref{f7} shows these for $N_a=5$ and $M=7$. We again see the inherited mirror symmetry with respect to the middle state.

%Figura 7

The density matrices of the eigenstates $k$ and $dim - k + 1$, where $k$ labels the states in order of increasing energy and $dim$ is the dimension of the Hilbert space of the system, have the same coefficients and their coherences differ only by a phase, except for the degenerate states $E=M$.

%Figura 8
\begin{figure}
\begin{center}
a)\ \includegraphics[width=1.0\linewidth]{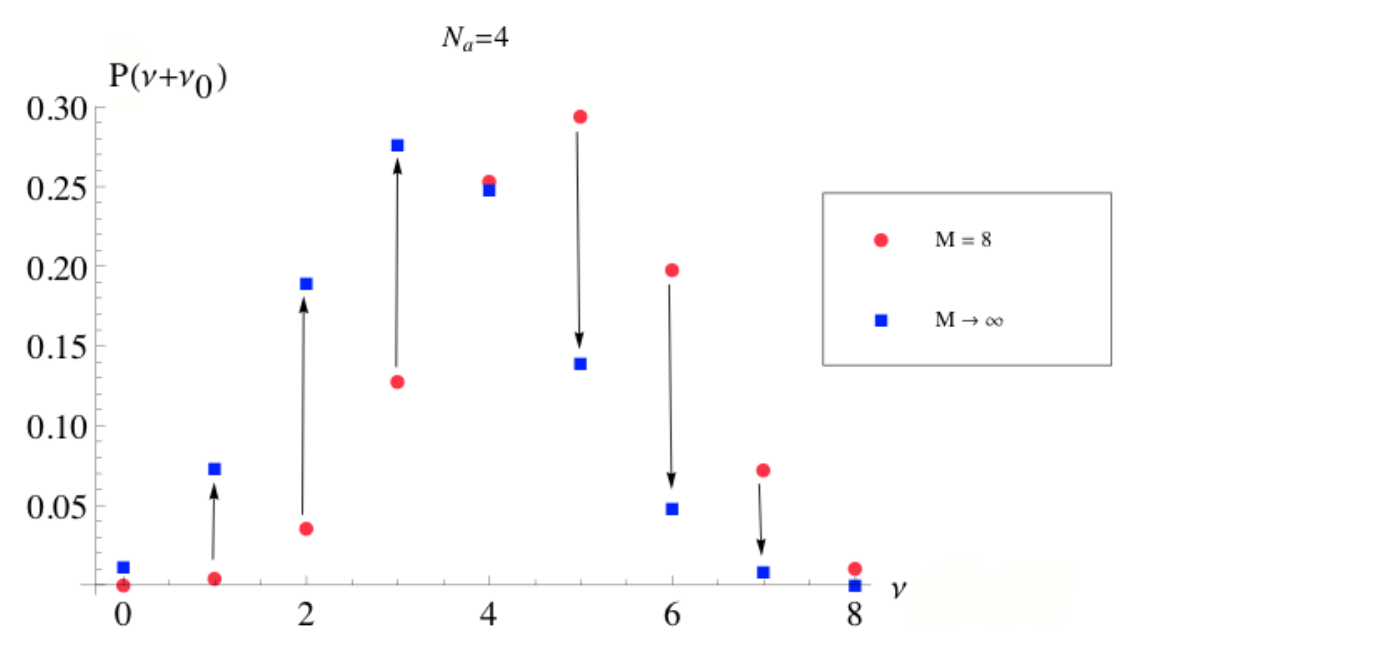} \\[1cm]
b)\ \includegraphics[width=1.0\linewidth]{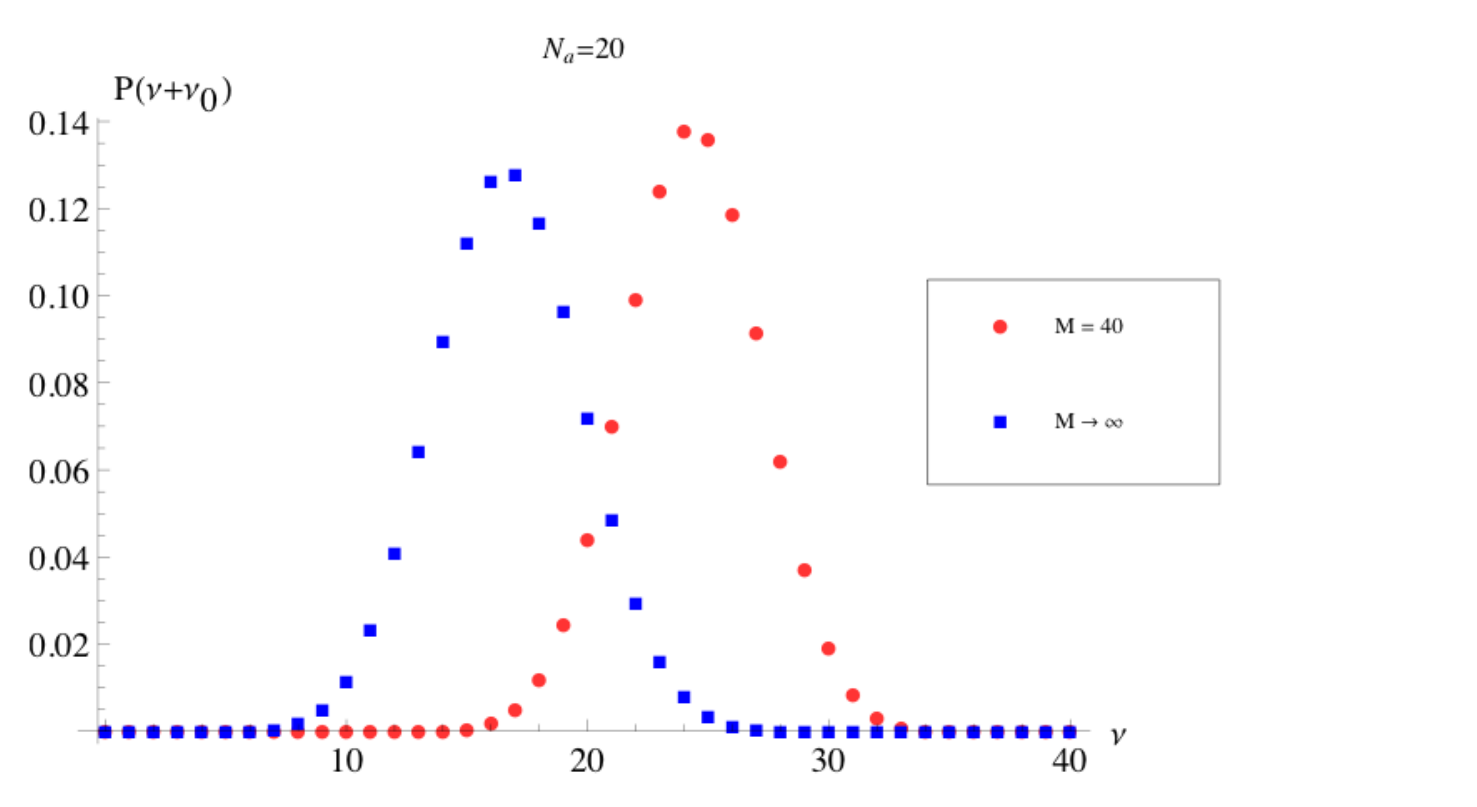} 
\end{center}
\caption{(Color online.)  Probability $P(\nu+\nu_0)$ of having $\nu+\nu_0$ photons in the system at the triple point in double resonance, as the excitation number $M$ grows without limit.}
\label{f8}
\end{figure}

Another interesting limit is $M \to \infty$, where the dimension of the Hilbert space depends only on $N_a$~\cite{qts8}. For a fixed number $N_a$ of atoms, as $M$ increases the atomic excitations may saturate and the population of photons grows. It is therefore interesting to study the photon distribution in the system at the triple point. The probability of having a certain number $\nu$ of photons is obtained by taking the modulus squared of the scalar product of the Fock state $\vert\nu\rangle$ and the eigenstate. Given $N_a$ and $M$, the minimum number of photons in the QED cavity is $\nu_0 = M-2\,N_a$. We consider here the ground state at each value of the excitation number $M$ and evaluate the excess number of photons, i.e., the probability $P(\nu + \nu_0)$ of having $\nu + \nu_0$ photons in the system. This is plotted in Figure~\ref{f8}a for $N_a=4$ atoms and for $M=8$ and the limit $M\to\infty$. The arrows in the figure show how each value moves as $M$ grows without limit. Both distributions are quasi-gaussian, though this is not very clear in the figure. For a larger number of atoms, $N_a=20$, Fig.~\ref{f8}b shows better the photon probability distributions, this time for $M=40$ and $M\to\infty$.

\section{Conclusions}

A triple point in the phase diagram of a three-level system interacting with a one-mode electromagnetic field is found for the atomic $\Xi$-configuration. This triple point is fixed in parameter space, is present for any finite number of atoms, and prevails in the thermodynamic limit. When in double resonance it resides at $(\mu_{12},\,\mu_{23}) = (1,\,\sqrt{2})$. Away from the double resonance condition it acquires different coordinates, but it is still fixed as $N_a\to\infty$. Properties of observables at the triple point were studied, including state structure, atomic level populations, energy spectra, and photon number probabilities. We find that the energy spectrum has a mirror symmetry with respect to the point $E=M$, where $M$ is the total excitation number (a constant of motion of the system).

\vspace{0.2in}

{\it Acknowledgements:} This work was partially supported by CONACyT-M\'exico (under project 101541) and DGAPA-UNAM (under projects IN101614 and IN110114).


\begin{thebibliography}{99}

\bibitem{yi}
X.X. Yi, X.H. Su and L. You, {\it Phys. Rev. Lett.} {\bf 90}, 097902 (2003).

\bibitem{jane}
E. Jan\'e, M.B. Plenio and D. Jonathan, {\it Phys. Rev.} A {\bf 65}, 050302® (2002)

\bibitem{hepp}
K. Hepp and E.H. Lieb, {\it Ann. Phys.} (NY) {\bf 76}, 360 (1973).

\bibitem{cordero1}
S. Cordero, R. L\'opez-Pe\~na, O. Casta\~nos and E. Nahmad-Achar, {\it Phys. Rev.} A {\bf 87}, 023805 (2013).

\bibitem{cordero2}
S. Cordero, O. Casta\~nos, R. L\'opez-Pe\~na and E. Nahmad-Achar, {\it J. Phys. A: Math. Theor.} {\bf 46}, 505302 (2013).

\bibitem{SimInNat}
O. Casta\~nos, E. Nahmad-Achar, R. L\'opez-Pe\~na and J.G. Hirsch, in {\it Symmetries in Nature} (L. Benet, P.O. Hess, J. M. Torres and K. B. Wolf, eds.), AIP Conf. Proc. {\bf 1323}, 40 (2010).

\bibitem{gu}
S-J Gu, {\it Int. J. Mod. Phys.} B {\bf 24}, 4371 (2010).

\bibitem{perez}
P. P\'erez-Fern\'andez et al., {\it Phys. Rev.} E {\bf 83}, 046208 (2011).

\bibitem{caprio}
M.A. Caprio, P. Cejnar and F. Iachello, {\it Ann. Phys.} (NY) {\bf 323}, 1106 (2008).

\bibitem{qts8}
O. Casta\~nos, S. Cordero, R. L\'opez-Pe\~na and E. Nahmad-Achar, {\it J. Phys. Conf. Ser.} (2014) (to be published).

\end{thebibliography}
\end{document}